\documentclass{elsart}
\usepackage{epsfig}
\usepackage{amssymb}
\journal{World Scientific}

\newcommand{\msun}{\mbox{$M_\odot$}}

\def\be{\begin{eqnarray}}
\def\ee{\end{eqnarray}}
\def\lsim{\mathrel{\rlap{\lower3pt\hbox{\hskip1pt$\sim$}}
     \raise1pt\hbox{$<$}}} 
\def\gsim{\mathrel{\rlap{\lower3pt\hbox{\hskip1pt$\sim$}}
     \raise1pt\hbox{$>$}}} 

\begin{document}

\runauthor{Bethe, Brown, \& Lee}

\begin{frontmatter}
\title{And Don't Forget The Black Holes\thanksref{tit}}
\thanks[tit]{
This paper was finished about a year before Hans Bethe died.
We submitted it to one of the leading popular scientific journals.
The editor we submitted it to returned it to us the next day saying
that there was insufficient interest to warrant publication.
Given the large number of letters now being published on short
$\gamma$-ray bursts and the fact that the merging of black-hole,
neutron-star binaries appear to be as good as, if not better,
than the merging of neutron-star, neutron-star binaries as a
source of these bursts, we believe our factor of $\sim 20$ in the
ratio of the former to the latter to be highly relevant. 
Therefore we put our paper in the archives. It will be published
in {\it Hans Bethe and his Physics} by World Scientific
(2 July, 2006).
}

\author[cornell]{H.A. Bethe\corauthref{bethe}}
\author[suny]{G.E. Brown}
\author[pnu]{C.-H. Lee}

\address[cornell]{Floyd R. Newman Laboratory of Nuclear Studies,\\
Cornell University, Ithaca, NY 14853}

\address[suny]{Department of Physics and Astronomy,
               State University of New York, \\ Stony Brook, NY 11794, USA}

\address[pnu]{
Department of Physics, 
Pusan National University,
              Busan 609-735, Korea\\ (E-mail: clee@pusan.ac.kr) }

\corauth[bethe]{Deceased}

\begin{abstract}
The discovery of the highly relativistic neutron star (NS) binary
(in which both NS's are pulsars) not only increases
the estimated merging rate for the two NS's by
a large factor, but also adds the missing link in the
double helium star model of binary NS evolution.
This model gives $\sim 20$ times more gravitational merging
of low-mass black-hole (LMBH), NS binaries than binary
NS's, whatever the rate for the latter is.
\end{abstract}

\end{frontmatter}



The recent discovery of Burgay et al. \cite{Burgay} 
of the double pulsar PSR J0737$-$3039A and
PSR J0737$-$3039B \cite{Lyne} is very interesting for many reasons.
One that is not so obvious is that it involves just about the
lowest possible mass main sequence giants as progenitors,
which have not been encountered in the evolution of the
other binary NS's, even though there are 20 to 30 times
more of them, as we show below.

The observational evidence is strong that
NS binaries evolve from double helium stars,
avoiding the common envelope evolution of the standard scenario \cite{vdHvP}.
(A helium star results in a giant when the hydrogen envelope is
lifted off - in binary evolution by being transferred to the 
less massive giant
in the binary.) In the standard scenario of binary NS formation
after the more massive giant transfers its hydrogen envelope to the
companion giant, the remaining helium star burns and then explodes into a
NS. In about half the cases the binary is not
disrupted in the explosion. The NS waits until the
remaining giant evolves (and expands)
in red giant following its main sequence hydrogen burning.
Once the envelope is close enough to the NS, the
latter couples to it hydrodynamically through gravity.
In the system in which the NS is at rest, it sees
the envelope matter coming at it.
Some of it is accreted onto the NS, although
most flies by in the wake, being heated in the process, and is lost
into space. The energy to expel the matter comes from the drop
in potential energy as the orbit of the NS tightens. 
Formulas for the
tightening and the amount of mass accreted by the NS 
were given by Bethe \& Brown  \cite{BB98}.

Chevalier \cite{Chevalier93} first estimated that in the common envelope
evolution the NS would accrete sufficient matter to evolve
into a black hole (BH). This was 
made quantitative by Bethe \& Brown \cite{BB98}
who calculated that in a typical case, the NS would accrete
$\sim 1\msun$, taking it into an $\sim 2.4\msun$ 
low mass black hole (LMBH),
similar to the BH we believe resulted from SN1987A.

Thus, if the NS had to go through common envelope evolution
in a hydrogen envelope of $\gsim 10\msun$ from the giant, it
would accrete sufficient matter to go into a LMBH.
Therefore, when the giant evolved into a helium star, which later
exploded into a NS, a LMBH-NS binary would result provided the
system was not broken up in the explosion. (About 50\% of
the time the system survives the explosion.)
The above scenario was estimated \cite{BB98}
to take place 10 times more frequently than binary NS
formation which required the two stars to burn helium at the same time,
and, because of the greater mass of the BH,
the mergings of binaries with LMBH to be twice as likely to be
seen as those with only NS's. This is the origin of the factor
20 enhancement of gravitational mergers to be observed at LIGO,
over the number from binary NS's alone.

We review the known
NS binaries and show that they are consistent with the two
NS's in a given binary having very nearly the same mass,
as would follow from the double helium star scenario.

\begin{table}
\caption{5 observed neutron star binaries with measured masses.}
\label{tab1}
\begin{tabular}{lllll}
\hline
Object      & Mass ($\msun$) & Object    & Mass ($\msun$) & 
Refs.\\
\hline
J1518$+$4904           & 1.56$^{+0.13}_{-0.44}$ &
J1518$+$4904 companion & 1.05$^{+0.45}_{-0.11}$ & \cite{Thorsett,Nice}\\
B1534$+$12           & 1.3332$^{+0.0010}_{-0.0010}$ &
B1534$+$12 companion & 1.3452$^{+0.0010}_{-0.0010}$ & \cite{S98}\\
B1913$+$16           & 1.4408$^{+0.0003}_{-0.0003}$ &
B1913$+$16 companion & 1.3873$^{+0.0003}_{-0.0003}$ & \cite{S99}\\
B2127$+$11C           & 1.349$^{+0.040}_{-0.040}$ &
B2127$+$11C companion & 1.363$^{+0.040}_{-0.040}$ & \cite{S111} \\
J0737$-$3039A        & 1.337$^{+0.005}_{-0.005}$ &
J0737$-$3039B        & 1.250$^{+0.005}_{-0.005}$ & \cite{Lyne} \\
\hline
\end{tabular}
\end{table}

We list in Table~\ref{tab1} the 5 observed NS binaries with measured
masses.
The very nearly equal masses of pulsar and companion in 1534$+$12
and 2127$+$11C is remarkable. We show below that 1913$+$16 comes from
a region of giant progenitors in which the masses could easily be
as different as they are. The uncertainties in 1518$+$49 are great
enough that the masses could well be equal. The main point in our
letter is to show that the masses in the double pulsar
J0737$-$3039A and J0737$-$3039B were probably very nearly the same
before a common envelope evolution in which the first formed
NS J0737$-$3039A accreted matter from the evolving
(expanding) helium star progenitor of J0737$-$3039B 
in the scenario of Dewi \& van den Heuvel \cite{Dewi}.

Since the helium burning in the giant is an order of magnitude
shorter than the hydrogen burning in time, for two giants to burn
helium at the same time they have to be very nearly equal in mass,
within $4\%$. Thus the binary neutron star scenario is highly
selective.
If they do burn helium at the same time, they can go
through common envelope evolution at this time.
The hydrogen envelope of the slightly more massive giant expands
in red giant and is transferred to the less massive giant,
which in turn evolves in red giant. The time for these evolutions
is so short that the helium stars are unable to accept the hydrogen
which is lost into space \cite{BraunLanger} so that
the final neutron stars are also close in mass.
Each of the helium stars will explode, going into a NS.
In about half of each explosion the binary is disrupted, but
in about 1/4 of the cases it is preserved and a double neutron
star results.

Bethe \& Pizzochero \cite{BP90} used for SN 1987A
a schematic but realistic treatment of the radiative transfer
problem which allowed them to follow the position in mass of the
photosphere as a function of time. They showed that the
observations determine uniquely the kinetic energy of the envelope
once the mass is known. From the envelope masses considered,
the range of energies was 1 to $1.4\times 10^{51}$ ergs. Using the fact
that the pressure following the supernova shock is radiation dominated,
Bethe \& Brown \cite{BB95} showed from the known value of $0.075\msun$ of 
$^{56}$Ni production, an upper limit on the gravitational mass of
$1.56\msun $ could be obtained. This is just the calculated Fe core
mass for the known $18\msun$ progenitor of 1987A.

Calculations of Woosley are shown in Table~3 of Brown et al.
\cite{BWW} where the amounts of fallback material from distance of
3500 km and 4500 km are given. The fallback simply cancels the
gravitational binding energy from the Fe core evolving into a NS.
Since the density of matter is thin at the estimated 4000 km
bifuraction point, this cancellation is relatively insensitive
to the precise point of bifurcation. Thus, we can use the Fe
core mass as the mass of the NS.

\begin{figure}
\centerline{\epsfig{file=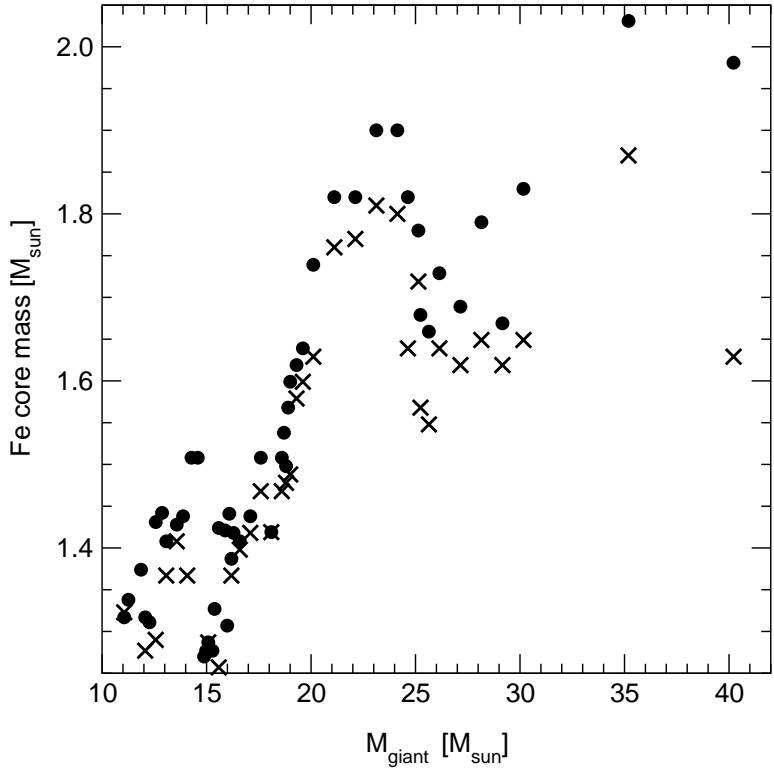,height=5in}}
\caption{Fe core masses for a grid of stellar masses. See text
for explanation.}
\label{fig1}
\end{figure}

In Fig.~\ref{fig1}, 
we show both recent and older calculations of Fe core
masses. The filled circles and crosses correspond to core masses 
at the time of iron core implosion for a finely spaced grid of stellar
masses \cite{Heger01}. The circles were calculated with the
Woosley \& Weaver code \cite{WW95}, whereas the crosses employ the
vastly improved rates for electron capture and
beta decay \cite{Langanke00}. 

A rapid increase in Fe core masses occurs at $M_{\rm giant}\sim 18\msun$,
just the mass of the progenitor of 1987A which we believe went into a
LMBH. The Burrows \& Woosley \cite{Burrows} evolution of 1913$+$16
took place in terms of $M_{\rm giant}\sim 20\msun$.
(In this evolution the Fe core mass will be slightly less than
given in Fig.~\ref{fig1} because the mass transfer in binary
evolution removes the hydrogen envelope, leaving a ``naked" helium star.
The wind loss from naked helium stars is great.) Now one can see
that the difference in masses between,
19 and $20\msun$ giants is greater than $0.1\msun$, so that the 
$0.05\msun$ difference in masses between the pulsar and its companion
in 1913$+$16 can easily be furnished out of the calculated Fe core
masses.

There is some tendency for the Fe core masses to decrease from
$M_{\rm giant}$ of 15 to $16\msun$, where we would estimate 1534$+$12
and its companion to come from. (The companion is slightly more
massive than the pulsar.)

The pulsar and companion in the double
pulsar should come from $M_{\rm giant}\sim 10\msun$, at the
$1.25\msun$ of the present J0737$-$3039A.
Following Dewi \& van den Heuvel \cite{Dewi}, the first pulsar
formed is increased in mass with mass transfer from the companion
when the latter is a helium star. Low-mass helium stars,
2.3 to $3.3\msun$ have to burn hotter than higher mass ones 
because of the greater energy loss
from their surfaces. In reaction to this, they expand in a helium
red giant phase.
The pulsar now goes through a common envelope
stage with the envelope of the helium star once the latter expands
far enough to make contact.

As noted, the NS accretes some of the matter and
most is expelled in the wake. Thus, from the coefficient
of dynamical friction
$c_d =2\ln (b_{\rm max}/b_{\rm min})\simeq 6$ \cite{Shima},
it is found \cite{BB98} that
\be
\frac{M_{\rm pulsar, f}}{M_{\rm pulsar, i}}
=\left(\frac{M_{\rm He,f}/a_{\rm f}}{M_{\rm He,i}/a_{\rm i}}
\right)^{1/5}.
\label{eq1}
\ee
We take $M_{\rm He,i}=2.5\msun$ and $M_{\rm He,f}=1.5\msun$.
Now the orbit in the double pulsar 
is decreased a factor of $\sim 3.7^{2/3}$ from that of
the binary NS. (The average of periods of 1913$+$16 and 1534$+$12
is $0.37$ days, whereas that of the double pulsar is 0.1 day.)
The $2/3$ exponent is the Kepler relation between orbital period and
radius, and we then find
\be
\frac{M_{\rm pulsar,f}}{M_{\rm pulsar,i}} = 1.075
\ee
which gives the correct $1.34\msun$ for J0737$-$3039B
given $M_{\rm pulsar, i}=1.25\msun$, the same as the unrecycled
neutron star.
The $1.5\msun$ He star will burn into the unrecycled pulsar,
with mass decreased $\sim 15\%$ by the general relativity
binding energy.

Therefore, in the double pulsar, the masses of the two giant progenitors
must be very close, since the two NS's would have had
nearly equal masses (as in 1534$+$12), except for the mass transfer
during the helium red giant stage.

The Salpeter mass distribution of giants gives the relative
number proportional to $(M_{\rm giant})^{-2.35}$, so for two
equal mass giants, the distribution goes as
$(M_{\rm giant})^{-4.7}$.
Thus, the probability of two $10\msun$ giants is 26 times greater
than that of two $20\msun$ giants, as would be progenitors of 1913$+$16.
Burgay et al. \cite{Burgay} estimated the ratio
$N_{0737}/N_{1913}$ to peak at $\sim 6$, whereas Dewi \& van den Heuvel
\cite{Dewi}
estimate that there are many of the PSR J0739-like systems
not seen because of their weak signals compared with those from
1913$+$16, possibly giving a factor $\sim 30$ in mergings.
Our estimate based on the large number of low-mass
giant progenitors confirms this.

With $\gsim 10\msun$ hydrogen envelopes and a proportional 
tightening of the orbit in the common envelope evolution in the
standard scenario, the Bethe \& Brown result of accretion of $\sim 1\msun$
onto the first NS seems reasonable.
After only part of this is accreted, the neutron star, like in
1987A, goes into a black hole.

Why haven't we seen any LMBH-NS binaries? Van den Heuvel \cite{vdH,vdH2}
has pointed out that NS's form with strong magnetic fields
$10^{12}$ to $5\times 10^{12}$ gauss, and spin down in a time
\be
\tau_{\rm sd} \sim 5\times 10^6 \; {\rm years}
\ee
and then disappear into the graveyard of NS's.
(The pulsation mechanism requires a minimum voltage from the polar
cap, which can be obtained from $B_{12}/P^2 \sim 0.2$
with $B_{12}=B/10^{12}{\rm G}$ and $P$ in seconds \cite{vdH2}.) 
The relativistic binary PSR 1913$+$16 has
a weaker field $B\simeq 2.5\times 10^{10}$ gauss and therefore
emits less energy in magnetic dipole radiation. Van den Heuvel estimates
its spin-down time as $10^8$ yrs. There is thus a premium in observational
time for lower magnetic fields, in that the pulsars can be seen for
longer times.

Taam and van den Heuvel \cite{Taam86} found empirically that the
magnetic field of a pulsar dropped roughly linearly with accreted
mass. This accretion can take place from the companion in any stage
of the evolution.
A pulsar that has undergone accretion is said to have
been ``recycled".

Now as mentioned, in 1913$+$16 the pulsar magnetic field is
$\sim 2.5\times 10^{10}$ gauss and in 1534$+$12 it is 
$\sim 10^{10}$ gauss. In J0737$-$3039A it is only
$6.3\times 10^9$ gauss. These ``recycled pulsars" will
be observable for $\gsim 100$ times longer than a ``fresh" (unrecycled)
pulsar. 

The same holds for LMBH-NS binaries. The NS is certainly not recycled,
so there is an about 1\% chance of seeing one as the recycled pulsar
in a binary NS. But we propose 10 times more of the LMBH's than binary
NS's, of which we observe 5. Thus the total probability of seeing
the LMBH binary should be about 50\%. 
However, there may be additional reasons that the
LMBH-NS binary is not observed. NS's in binaries in which the
nature of the companion star is unknown are observed. 
We may have to wait for
LIGO which will be able to measure ``chirp" masses quite accurately.
The chirp mass of a NS binary should concentrate near $1.2\msun$,
whereas the LMBH-NS systems should have a chirp mass
of $1.6\msun$, and there should be $\sim 20$ times more of the
latter.

In order to estimate the number of mergings per year for the LIGO
we begin from the estimate of Kim et al. \cite{Kim}
and multiply by a factor of 26 to account for the increase in this
estimate by the large number of low-mass giant progenitors,
neutron stars from two of these having been observed in the double
pulsar. Additionally we multiply by the factor of 20
to include NS-LMBH binary mergings. This brings our estimate
up to 0.5 to 3.6 mergings per year for the initial LIGO,
and about 6000 times greater for the advanced LIGO.



\section*{Acknowledgments}
G.E.B. was supported in part by the
US Department of Energy under Grant No. DE-FG02-88ER40388.
C.H.L. was supported by the Korea Research Foundation
Grant funded by the Korea Government (MOEHRD)(KRF-2005-070-C00034).

\end{document}